\documentclass[floats,aps,onecolumn,showpacs]{revtex4}
\usepackage{dcolumn}
\usepackage{epsfig}

\newcommand {\beq} {\begin{eqnarray}}
\newcommand {\eeqn} [1] {\label{#1} \end{eqnarray}}%

\begin{document}
\title{Convergence of the self-energy in a relativistic chiral quark model:
   excited Nucleon and  $\Delta$ sector}
\author{
E.\ M.\ Tursunov
}
\affiliation{
 Institute of Nuclear Physics, Uzbekistan
Academy of Sciences, 100214, Ulugbek, Tashkent, Uzbekistan
 }

\date{\today}
\begin{abstract}
\par A convergence of the valence quark self-energies in the
$1S$, $2S$, $1P_{1/2}$, $1P_{3/2}$ orbits induced by pion and gluon
field configurations, is shown in the frame of a relativistic chiral
quark model. It is shown that in order to reach a convergence, one
needs to include contribution of the intermediate quark and
anti-quark states with the total momentum up to $j=25/2$. It is
argued that a restriction to the lowest mode when estimating the
self-energy is not good approximation.
\end{abstract}


\maketitle

\section{Introduction}
\par Current programs  of the world leading facilities such as JLAB
and MAMI, aims to study the properties of the excited $N^*$ and
$\Delta^*$ resonances (see Ref. \cite{diaz1,diaz2,jan1,nstar07}).
New data obtained in these experiments yield valuable information on
effective degrees of freedom in the low-energy regime of the QCD.
\par In spite of many efforts, there still exist a number of serious
problems in the baryon spectroscopy. First of them is the problem of
missing resonances, which were predicted by the Constituent Quark
Models (CQM) \cite{cap86,glo98, met01}, but are still being not
observed at the current experiments. Additionally, the structure of
the Roper resonance is not yet well understood. It is not yet clear,
what is the mechanism which lowers the Roper's mass.
\par An opposite picture on the baryon structure is based on the
relativistic description. Although the CQMs are well adapted for the
calculations of the ground and excited baryon spectrum, arguments
against the CQM are very serious: a valence quark inside the baryon
moves with a velocity, close to the speed of the light. Another fact
is that a single pion or gluon can be emitted by a valence quark and
be absorbed  by the same  quark (through self-energy diagrams).
After emitting a single boson, the intermediate quark can excite to
the quark or antiquark states, thus yielding the contribution of the
sea-quarks to the baryon mass-spectrum. Of course, such diagrams can
not be involved in the CQMs. An alternative relativistic Cloudy Bag
Model \cite{thomas80} was not extended to the excited Nucleon and
$\Delta$ sector, though it could explain the proton spin problem
\cite{thomas08}.
\par In Ref. \cite{tur05,tur09} we have developed a relativistic
chiral quark model \cite{ oset84,gutsche87,gutsche89} for the
excited $N^*$ and $\Delta^*$  spectroscopy. The model includes
one-gluon and one-pion loop diagrams contributions and reproduces
the main properties of the lowest excitation spectrum of the Nucleon
and $\Delta$. It was shown that relativistic one-meson and one-gluon
exchange forces  generate spin 0, 1, 2, etc. operators, which couple
the lower and the upper components of the two interacting valence
quarks. The only contribution to the ground state nucleon and
$\Delta$ comes from the spin 1 operators, which correspond to the
exchanged pion or gluon in the l=1 orbit. This result can explain
naturally, why the $\Delta (3/2^+) (1232)$ resonance appears in the
$P$-wave of the $\pi N$ scattering data, which can not be understood
in the frame of the Constituent Quark Models.
\par The aim of present paper is to show the convergence of the valence quark
self-energy diagrams induced by the pion and gluon field
configurations. A convergence of the self-energy induced by pion
fields in the present model has been demonstrated for the valence
quarks in the ground state \cite{gutsche87,gutsche89}. However, an
extension of the model to the excited $N^*$ and $\Delta^* $ requires
to study the behavior of the self-energy for the excited valence
quark states. As it is well known \cite{saito84}, the convergence of
the self-energy is a serious problem for the bag models due-to the
sharp surface quark-pion coupling. It will be shown in our model,
that to reach convergence, one needs to include the contribution of
the intermediate quark and antiquark states with a total momentum up
to $j=25/2$. This is why a restriction to the lowest modes used in
many relativistic calculations is not a good approximation
\cite{saito08}.
\par The relativistic quark model is based on an effective chiral Lagrangian
describing quarks as relativistic fermions moving in a confining
static potential. The potential is described by a Lorentz scalar and
the time component of a vector potential, where the latter term is
responsible for short-range fluctuations of the gluon field
configurations \cite{lus81}. The model potential defines unperturbed
wave functions of the quarks which are subsequently used in the
calculations of baryon properties. The baryons are considered as
bound states of valence quarks, surrounded by a pion cloud as
required by the chiral symmetry and by gluons. Interaction of quarks
with a pion is introduced on the basis of the linearized
$\sigma$-model \cite{gell60,gutsche87,gutsche89}. The quark-gluon
interaction is introduced on the field-theoretical basis as
prescribed by QCD. Calculations are performed perturbatively to
second order in the quark-pion and quark-gluon interaction. All
calculations are performed at one loop or at order of accuracy
$o(1/f_{\pi}^2, \alpha_s)$.

\par In the following we first give the basic formalism of our
approach, then present the numerical results.

\section{Model}
\par The effective Lagrangian of our model
${\cal L}(x)$ contains the quark core part ${\cal L}_Q(x)$ the quark-pion
 ${\cal L}_I^{(q\pi)}(x)$ and the quark-gluon ${\cal L}_I^{(qg)}(x) $
 interaction parts, and the kinetic parts for the pion ${\cal L}_{\pi}(x)$ and
gluon ${\cal L}_{g}(x)$:
\begin{eqnarray}
\nonumber
{\cal L}(x) = {\cal L}_Q(x) + {\cal L}_I^{(q\pi)}(x) + {\cal
  L}_I^{(qg)}(x)+ {\cal L}_{\pi}(x) +  {\cal L}_{g}(x)  \\
 \nonumber
 = \bar\psi(x)[i\not\!\partial -S(r)-\gamma^0V(r)]\psi(x) - 1/f_{\pi}
 \bar\psi[S(r) i \gamma^5 \tau^i \phi_i]\psi- \\
 -g_s \bar\psi A_{\mu}^a\gamma^{\mu}\frac{\lambda^a}{2} \psi +
  \frac{1}{2}(\!\partial_{\mu}\phi_i)^2 -{1\over 4}
G^a_{\mu\nu} G_a^{\mu\nu}.
\end{eqnarray}
Here, $\psi(x)$, $\phi_i, i=1,2,3$ and $A_{\mu}^a$ are the quark,
the pion and the gluon fields, respectively. The matrices $\tau^i
(i=1,2,3)$ and $\lambda^a (a=1,...,8)$ are the isospin and color
matrices, correspondingly. The pion decay constant $f_\pi=$93 MeV.
The scalar part of the static confinement potential is given by
\begin{equation}
S(r)=cr+m
\end{equation}
where c and m are constants.

\par At short distances, transverse fluctuations of the string are dominating
\cite{lus81}, with some indication that they transform like the time component of the
Lorentz vector. They are given by a Coulomb type vector potential as
\begin{equation}
\label{Coulomb}
 V(r)=-\alpha/r
\end{equation}
where $\alpha$ is approximated by a constant.
The quark fields are obtained from solving the Dirac equation with the corresponding
scalar plus vector potentials
\begin{equation} \label{Dirac}
[i\gamma^{\mu}\partial_{\mu} -S(r)-\gamma^0V(r)]\psi(x)=0
\end{equation}
The respective positive and negative energy eigenstates as solutions to the Dirac
equation with a spherically symmetric mean field, are given in a general form as
\begin{eqnarray} \label{Gaussian_Ansatz}
 u_{\alpha}(x) \, = \,
\left(
\begin{array}{c}
g^+_{N\kappa }(r) \\
-i f^+_{N\kappa }(r) \,\vec{\sigma}\hat{\vec x} \\
\end{array}
\right)
\, {\cal Y}_{\kappa}^{m_j}(\hat{\vec x}) \,\chi_{m_t} \, \chi_{m_c} \, exp(-iE_{\alpha}t)
\end{eqnarray}

\begin{eqnarray}
 v_{\beta}(x) \, = \,
\left(
\begin{array}{c}
g^-_{N\kappa}(r) \\
-i f^-_{N\kappa}(r) \,\vec{\sigma}\hat{\vec x} \\
\end{array}
\right)
\, {\cal Y}_{\kappa}^{m_j}(\hat{\vec x}) \,\chi_{m_t} \, \chi_{m_c} \, exp(+iE_{\beta}t)
\end{eqnarray}
The quark and anti-quark eigenstates $u$ and $v$ are labeled by the
radial, angular, azimuthal, isospin and color quantum numbers $N,\,
\kappa,\, m_j,\, m_t$ and $m_c$, which are collectively denoted by
$\alpha$ and $\beta$, respectively. The spin-angular part of the
quark field operators
\begin{equation}
{\cal Y}_{\kappa}^{m_j}(\hat{\vec x})\,=\,[Y_l(\hat{\vec x})\otimes
\chi_{1/2}]_{jm_j} \, \, j=|\kappa|-1/2.
\end{equation}
The quark fields $\psi$ are expanded over the basis of positive and negative energy
eigenstates as
\begin{equation}
\psi(x)=\sum \limits_{\alpha} u_{\alpha}(x)b_{\alpha} +\sum \limits_{\beta} v_{\beta}(x)d^{\dag}_{\beta} .
\end{equation}
The expansion coefficients $b_{\alpha}$ and $d^{\dag}_{\beta}$ are
operators, which annihilate a quark and create an anti-quark in the
orbits $\alpha$ and $\beta$, respectively.
\par The free pion field operator is expanded over plane wave solutions as
\begin{equation}
\phi_j(x)=(2\pi)^{-3/2}\, \int\frac{d^3k}{(2\omega_k)^{1/2}}[a_{j{\bf k}}exp(-ikx)+a^{\dag}_{j{\bf k}}exp(ikx)]
\end{equation}
with the usual destruction and creation operators $a_{j{\bf k}}$ and
$a^{\dag}_{j{\bf k}}$ respectively. The pion energy is defined as \\
$\omega_k \,=\, \sqrt{k^2+m_{\pi}^2}. $ The expansion of the free
zero mass gluon field operators is of the same form.
\par In denoting the three-quark vacuum state by $ |0> $, the corresponding
noninteracting many-body quark Green's function (propagator) is given by
the customary vacuum Feynman propagator for a binding potential \cite{fet71}:
\begin{equation}
iG(x,x')\,=\, iG^F(x,x')\,=\,<0|T\{\psi(x) \bar\psi(x')\}|0>\,=\,
\sum \limits_{\alpha} u_{\alpha}(x)\bar u_{\alpha}(x')\theta(t-t') +
\sum \limits_{\beta} v_{\beta}(x)\bar v_{\beta}(x')\theta(t'-t)
\end{equation}
Since the three-quark vacuum state $|0>$ does not contain any pion
or gluon, the pion and gluon Green's functions are given by the
usual free Feynman propagator for a boson field:
\begin{equation}
i\Delta_{ij}(x-x')\,=\, <0|T\{\phi_i(x) \bar\phi_j(x')\}|0>\,=\,
i\delta_{ij}\int\frac{d^4k}{(2\pi)^4}\frac{1}{k^2-m_{\pi}^2+i\epsilon}
\,exp[-ik(x-x')] \, ,
\end{equation}

\begin{equation}
i\Delta^{(\mu\nu)}_{ab}(x-x')\,=\, <0|T\{ A^a_\mu (x)
A^b_{\nu}(x')\}|0>\,=\,
 i\delta_{ab}g^{\mu \nu}\int\frac{d^4k}{(2\pi)^4}\frac{1}{k^2+i\epsilon}
\,exp[-ik(x-x')] \,,
\end{equation}
(in the Coulomb gauge), where we choose $g^{\mu\nu}=\delta_{\mu\nu}
g^{\mu\mu}$, $g^{00}=-g^{11}=-g^{22}=-g^{33}=1$ .
\par Using the effective Lagrangian and the time-ordered perturbation theory
 one can develop a calculation scheme for the lower
excitation spectrum of the nucleon and delta. In the model the quark
core result ($E_Q$) is obtained by solving Eq.(\ref{Dirac}) for the
single quark system numerically. Since we work in the independent
particle model and limited with the lower excitation spectrum of the
nucleon and Delta, the bare three-quark state of the $SU(2)$-flavor
baryons corresponds to the  structure $(1S_{1/2})^2(nlj)$ in the
non-relativistic spectroscopic notation. The corresponding quark
core energy is evaluated as the sum of single quark energies with:
$$ E_Q=2E(1S_{1/2}) + E(nlj)$$ with an appropriate correction on the
center of mass motion \cite{tur09}.

 \par The second order perturbative corrections to the energy spectrum of the
 SU(2) baryons due to the pion field ($\Delta E^{(\pi)}$) and the gluon fields ( $\Delta
 E^{(g)}$) are calculated on the basis of the Gell-Mann and Low theorem :
 \begin{eqnarray}\label{Energy_shift}
\hspace*{-.8cm}
\Delta E=<\Phi_0| \, \sum\limits_{i=1}^{\infty} \frac{(-i)^n}{n!} \,
\int \, i\delta(t_1) \, d^4x_1 \ldots d^4x_n \,
T[{\cal H}_I(x_1) \ldots {\cal H}_I(x_n)] \, |\Phi_0>_{c}
\end{eqnarray}
with $n=2$, where the relevant quark-pion and quark-gluon interaction
Hamiltonian densities are
\begin{eqnarray}
{\cal H}_I^{(q\pi)}(x)= \frac{i}{f_{\pi}}\bar\psi(x)\gamma^5
\vec\tau\vec\phi(x)S(r)\psi(x),
\end{eqnarray}
\begin{eqnarray}
{\cal H}_I^{(qg)}(x)= g_s\bar\psi(x)A_{\mu}^a(x)\gamma^{\mu}
\frac{\lambda^a}{2}\psi(x)
\end{eqnarray}
The stationary bare three-quark state $|\Phi_0>$ is constructed from
the vacuum state using the usual creation operators:
\begin{equation}
|\Phi_0>_{\alpha\beta\gamma}=b_{\alpha}^+b_{\beta}^+b_{\gamma}^+|0>,
\end{equation}
where $\alpha, \beta$ and $ \gamma$ represent the quantum numbers of
the single quark states, which are coupled to the respective baryon
configuration. The energy shift of Eq.(\ref{Energy_shift}) is
evaluated up to second order in the quark-pion and quark-gluon
interaction, and generates self-energy and exchange diagrams
contributions.

\par The self-energy terms contain contribution both from
intermediate quark $(E>0)$ and anti-quark $(E<0)$ states. These
diagrams correspond to the case when a pion or gluon is emitted and
absorbed by the same valence quark which is excited to the
intermediate quark and anti-quark states. In this way one can
estimate the contribution of the sea-quarks to the hadron spectrum
that can not be done in non-relativistic quark models.
\par The pion part of the self energy term (pion cloud contribution)
 is evaluated as
\begin{eqnarray}
\Delta
E_{s.e.}^{(\pi)}=-\frac{1}{2f_{\pi}^2}\sum\limits_{a=1}^{3}\sum\limits_{\alpha
  ' \leq \alpha_F} \int
\frac{d^3\vec p}{(2\pi)^3p_0} \biggl\{ \sum\limits_{\alpha}\frac{V_{\alpha
      \alpha ' }^{a+}(\vec p)V_{\alpha \alpha ' }^{a}(\vec
        p)}{E_{\alpha}-E_{\alpha '}+p_0}-
 \sum\limits_{\beta}\frac{V_{\beta \alpha '}^{a+}(\vec p) V_{\beta
 \alpha '}^{a}(\vec p)}{E_{\beta}+E_{\alpha '}+p_0}\biggr\},
\end{eqnarray}
with $ p_0^2=\vec p^2 + m_{\pi}^2$. The $q-q-\pi$ transition form
factors are defined as:
\begin{eqnarray}
V_{\alpha\alpha'}^a(\vec p)=\int d^3 x \bar u_{\alpha}(\vec x)\Gamma^a(\vec x)
u_{\alpha '}(\vec x) e^{-i\vec p \vec x} \\
V_{\beta \alpha'}^a(\vec p)=\int d^3 x \bar v_{\beta}(\vec x)\Gamma^a(\vec x)
u_{\alpha '}(\vec x) e^{-i\vec p \vec x}
\end{eqnarray}
The vertex function of the $\pi -q -q $ and $\pi -q -\bar q $
transition is
\begin{eqnarray}
\Gamma^a= S(r) \gamma^5 \tau^a I_c \, ,
\end{eqnarray}
where $I_c$ is the color unity matrix.

\begin{figure}[tbh]
\begin{center}
\includegraphics[width=15cm]{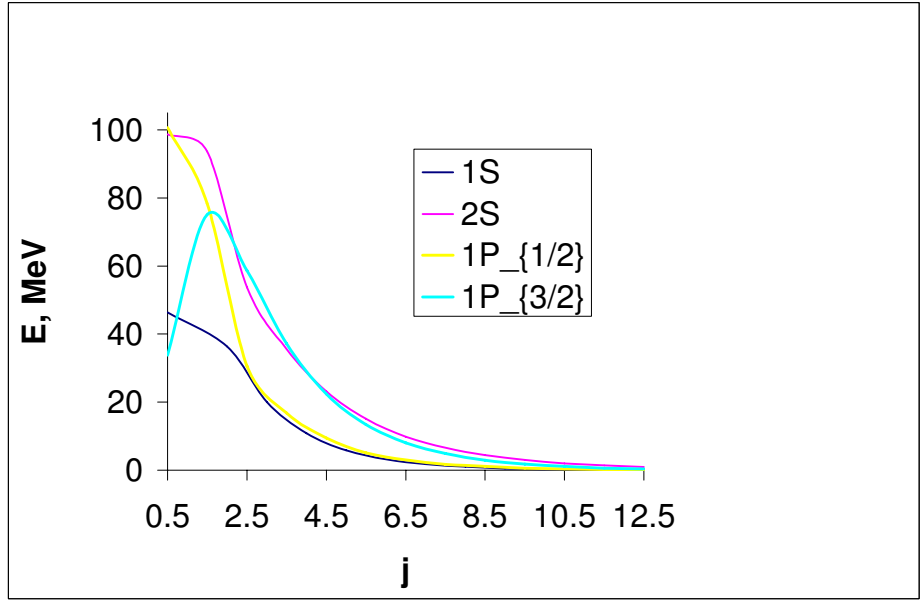}
\end{center}
\caption{Contributions self energy diagrams induced by $\pi-$meson
fields to the valence quark states in dependence on the total
momentum of the intermediate quark (antiquark) states \label{Fig1}}
\end{figure}

\par The gluon part of the second order self-energy diagrams (gluon cloud) contribution
is  estimated in a similar fashion as
\begin{eqnarray}
\Delta E_{s.e.}^{(g)}=\frac{g_s^2}{2}\sum\limits_{a}g_{\mu\nu}
\sum\limits_{\alpha ' \leq \alpha_F} \int\frac{d^3\vec p}{(2\pi)^3p}
\biggl\{ \sum\limits_{\alpha}\frac{V_{\alpha
      \alpha ' }^{a\mu+}(\vec p)V_{\alpha \alpha ' }^{a\nu}(\vec p)}
{E_{\alpha}-E_{\alpha '}+p}-
 \sum\limits_{\beta}\frac{V_{\beta \alpha '}^{a\mu+}(\vec p) V_{\beta
 \alpha '}^{a\nu}(\vec p)}{E_{\beta}+E_{\alpha '}+p}\biggr\},
\end{eqnarray}
where the transition form factor is evaluated with the corresponding
vertex matrix
\begin{eqnarray}
\Gamma_{\mu}^a=\gamma^{\mu}\frac{\lambda^a}{2}I_t
\end{eqnarray}
with the isospin unity matrix $I_t$.
\par  After evaluation of the
transition form-factors and integration over angular variables, the
self-energy terms induced by the both pion and gluon fields can be
estimated in a proper way (see Ref.\cite{tur09} for details).

 \begin{figure}[tbh]
\begin{center}
\includegraphics[width=15cm]{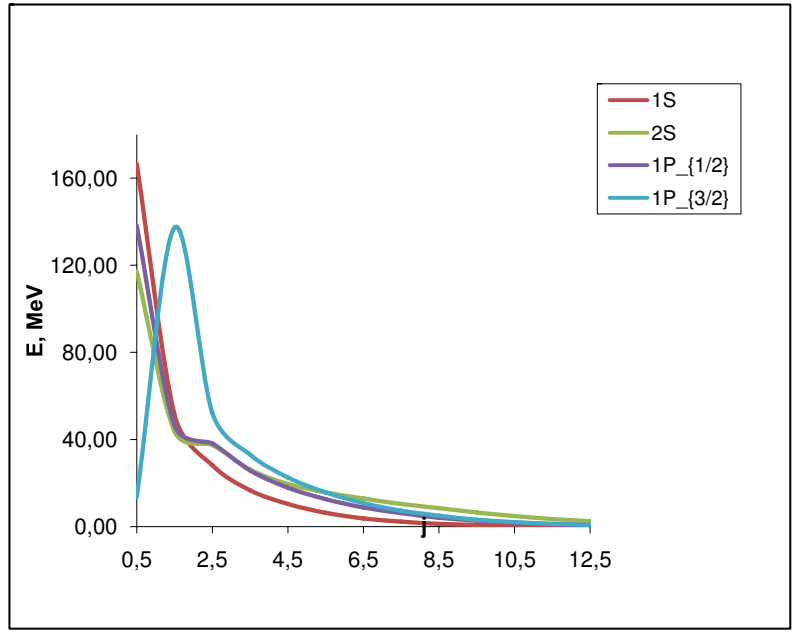}
\end{center}
\caption{Contributions self energy diagrams induced by
 by the color-electric components of the gluon fields
 \label{Fig2}}
\end{figure}

\begin{figure}[tbh]
\begin{center}
\includegraphics[width=15cm]{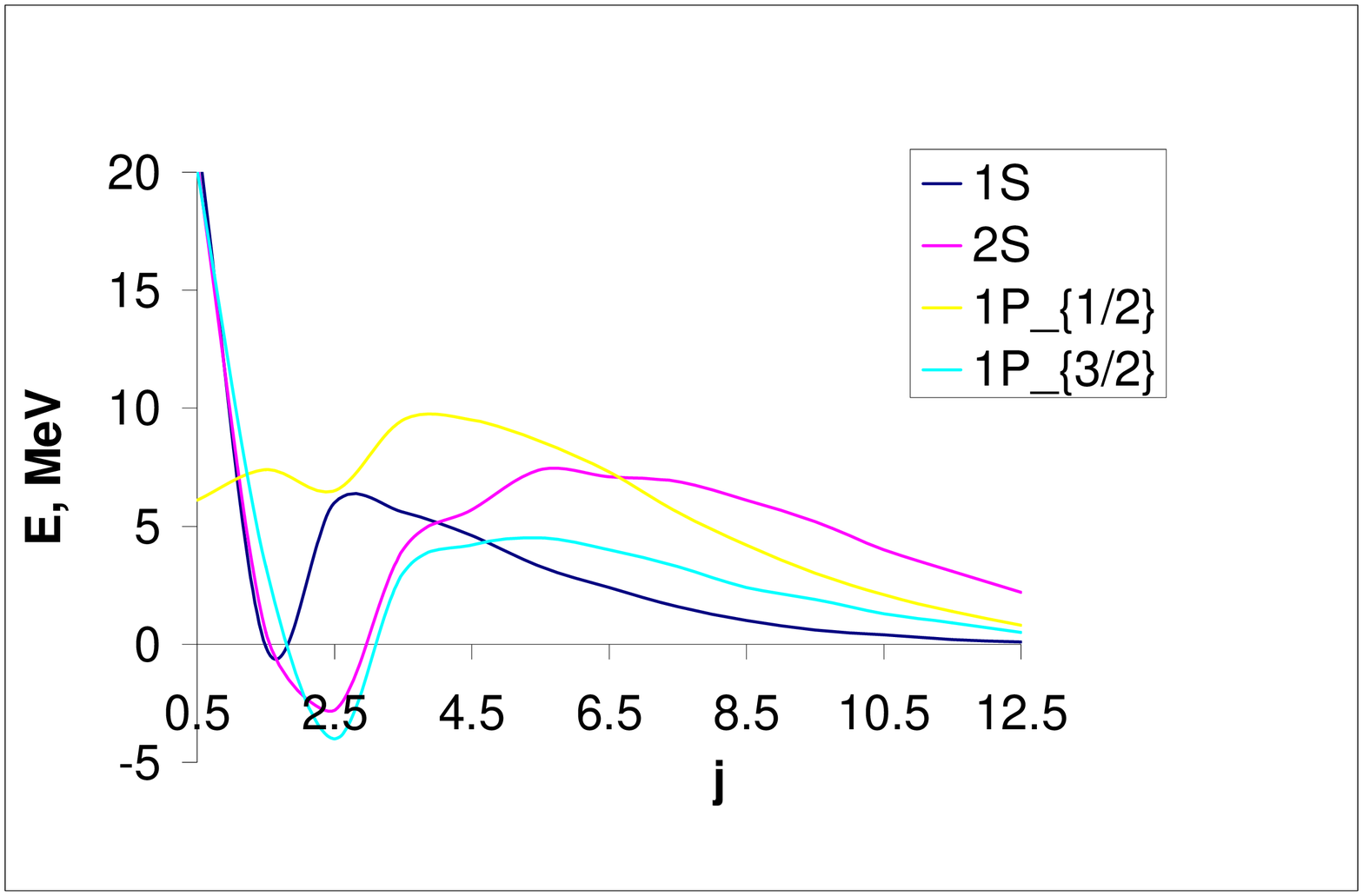}
\end{center}
\caption{Contributions self energy diagrams induced by
 by the color-magnetic components of the gluon fields
 \label{Fig3}}
\end{figure}

\section{Numerical results}

\par In order to account for the finite size effect of the pion, we introduce
an one-pion vertex regularization function in the momentum space,
parameterized in the dipole form as
$$ F_{\pi}(p^2)= \frac {\Lambda_{\pi}^2-m_{\pi}^2}{\Lambda_{\pi}^2+p^2}.$$
From the flux-tube study \cite{lus81} we fix $\alpha=0.26 \approx
\pi/12$. The strong coupling constant $g_s^2=4\pi\alpha_s$ with the
value $\alpha_s=0.65$. The parameters of the confining potential
($c=0.16$ $GeV^2$ and  $m=0.06$ $GeV$) are chosen to reproduce the
correct axial charge $g_A$ of the proton (and the empirical
pion-nucleon coupling constant $G^2_{\pi NN}/4\pi=14$ via the
 Goldberger-Treiman relation) and a normal value for the quark core RMS
 radius of the proton (see \cite{gutsche87}). The pion-form factor
 $\Lambda_{\pi}=1.0$ GeV (see Model A in Ref \cite{tur09}).
It was shown in the Ref. \cite{gutsche87} that the static properties
of the proton with the correction on the center of mass motion (CM)
can be well reproduced with above parameters of the model.

\par In Fig. \ref{Fig1} we present the self-energy of the single
valence quarks in the $1S$, $2S$, $1P_{1/2}$, $1P_{3/2}$ orbits
induced by the pion fields in the dependence on the total momentum
of the intermediate quark and anti-quark states from $j=1/2$ up to
$j=25/2$. One can note that the convergence is faster for the
valence quark in the ground state $1S$. The convergence also is
reached for other valence quark states. We note here that a
contribution of the intermediate quark states is negative, while
intermediate anti-quark states yield larger and positive
self-energy. As can be seen from the picture, the maximal
contribution comes from the intermediate states with $j=1/2$ for the
valence quarks except the $1P_{3/2}$ orbit, which gets the most
contribution from the intermediate quark and anti-quark states with
the total momentum $j=3/2$.  From this picture one can conclude that
the sum of the self energy terms is limited and reasonable
\cite{tur09}.
\par A very interesting behavior of the self-energy terms due-to
color-electric field configurations can be found in the
Fig.\ref{Fig2}. The self-energies of the  $1P_{1/2}$  and $2S$
valence quarks, after rapid decreasing, have a plat around the
interval $j=(3/2,5/2)$. Again, as in previous figures, we can see a
reasonable convergence at $j=25/2$ for all valence quark states.
 \par Finally, in Fig. \ref{Fig3} we show the behavior of the
 self-energy of the $1S$, $2S$, $1P_{1/2}$, $1P_{3/2}$ valence quarks
 due-to color-magnetic fields. It's shape is quite different from
 that shown in the Fig. \ref{Fig1} and Fig. \ref{Fig2}. For all the
 valence quark states, the self-energy has a minimum at small total momentum
 $j$ value. Even more, the intermediate quark states  (with a negative self-energy
  contributions) dominate at this point over the intermediate
  anti-quark states for the valence quarks except the $1P_{1/2}$
  state. By further increasing the total momentum $j$ up to 25/2, one can reach
  convergent results.
\par In Refs. \cite{tur05, tur09} it was shown, that the
restriction of intermediate quark to the lowest mode is not a good
approximation. As we see above, in order to reach convergent
results, one needs to go up to large values of the total momentum
$j$. The summary  contribution of the pion fields to the $N(939)$ is
about 200 MeV, while the gluon fields yield more than 500 MeV shift
with the strong coupling constant $\alpha_s=0.65$ mostly due-to
color-electric field configurations. Contrary, the restriction to
the intermediate ground state when estimating the self-energy (I=0)
yields too small values for the nucleon energy (about 500 MeV with
the appropriate correction on the center of mass motion). Such
estimations were obtained also for other lowest Nucleon and $\Delta$
states.

\section{Summary and conclusions}
\par We have shown a convergence of the $1S$, $2S$, $1P_{1/2}$, $1P_{3/2}$
valence quarks self-energies in the frame of a relativistic chiral
quark model. It was shown that the convergent results for the
self-energy due-to gluon and pion field configurations can be
obtained with the inclusion of the intermediate quark and anti-quark
states with the total momentum up to $j=25/2$. It is argued that the
restriction to the lowest mode when estimating the self-energy is
not a good approximation.

{\it Acknowledgements}. Author thanks  Th. Gutsche for useful
discussions.

\end{document}